\documentclass[twocolumn,showpacs,aps,prl,superscriptaddress,letterpaper]{revtex4}

\usepackage{graphicx}

\usepackage{epsfig}

\usepackage{amsmath}

\usepackage{amssymb}

\usepackage{lineno}

\usepackage{hhline}

\usepackage{multirow}

\usepackage{fancyheadings}

\usepackage{ifthen}

\usepackage{lscape}

\usepackage{rotating}

\RequirePackage{xspace}

\newcommand{\BABARPubYear}    {03}
\newcommand{\BABARPubNumber}  {034}

\newcommand{\SLACPubNumber} {10380}

\newcommand{\vvsmallrule}{\rule[-1.0mm]{0.0cm}{0.6cm}}
\def\psm {\ensuremath{p^*_{\rm min}}}

\input babarsym

\def\B      {\ensuremath{B}\hbox{ }}

\newcommand {\Bxlnu}{\ensuremath{\Bb \rightarrow X \ell^- \bar{\nu}}}
\newcommand {\Bxclnu}{\ensuremath{\Bb \rightarrow X_c \ell^- \bar{\nu}}}
\newcommand {\Bxulnu}{\ensuremath{\Bb \rightarrow X_u \ell^- \bar{\nu}}}

\newcommand {\mmx} {\ensuremath{\langle M_X \rangle}}
\newcommand {\mmxs} {\ensuremath{\langle M_X^2 \rangle}}
\newcommand {\mmxc} {\ensuremath{\langle M_X^3 \rangle}}
\newcommand {\mmxq} {\ensuremath{\langle M_X^4 \rangle}}
\newcommand {\mmxn} {\ensuremath{\langle M_X^n \rangle}}

\newcommand {\breco}{\ensuremath{B_{\rm reco}}}

\newcommand{\beq}{\begin{equation}}
\newcommand{\beqa}{\begin{eqnarray}}
\newcommand{\beqn}{\begin{eqnarray}}
\newcommand{\eeq}{\end{equation}}
\newcommand{\eeqa}{\end{eqnarray}}
\newcommand{\eeqn}{\end{eqnarray}}
\makeatletter
\def\slash#1{{\mathpalette\c@ncel{#1}}} 
\makeatother

\def\isprl{1}

\long\def\inst#1{\par\nobreak\kern 4pt\nobreak
    {\it #1}\par\vskip 10pt plus 3pt minus 3pt}

\def\varia1c-a{666}

\begin{document}

\begin{flushleft}
\babar-PUB-\BABARPubYear/\BABARPubNumber\\
SLAC-PUB-\SLACPubNumber\\
\end{flushleft}

\title[Short Title] {Measurements of Moments of the Hadronic Mass Distribution in \\
Semileptonic \boldmath$B$  Decays }

%
\author{B.~Aubert}
\author{R.~Barate}
\author{D.~Boutigny}
\author{F.~Couderc}
\author{J.-M.~Gaillard}
\author{A.~Hicheur}
\author{Y.~Karyotakis}
\author{J.~P.~Lees}
\author{V.~Tisserand}
\author{A.~Zghiche}
\affiliation{Laboratoire de Physique des Particules, F-74941 Annecy-le-Vieux, France }
% ...
\collaboration{The \babar\ Collaboration}
\noaffiliation

\begin{abstract}

We report a measurement of the first four moments of the
hadronic mass distribution in $\Bxclnu$ decays.
The measurements are based on 89 million $\FourS \to \BB$ events where the hadronic decay of one of the $B$ mesons is fully reconstructed and a charged lepton from the 
decay of the other $B$ meson is identified.
The moments are presented for minimum lepton momenta ranging from 0.9 to 1.6 \gev in the $B$ rest frame.
It is expected that such measurements will lead to improved determinations of \Vcb and \Vub.

\end{abstract}


\date{March 18, 2004}

\maketitle


\ifnum\isprl<1
\section{Introduction}
\label{sec:introduction}
\fi

In this paper 
we report measurements of the first four moments \mmxn, with $n=1,\ldots4$,
of the hadronic mass distributions in $\Bb\to X_c \ell^-\nub$ decays~\cite{chargecon}.
The moments are presented as a function of \psm, the lower limit on the  charged lepton momentum, which we vary between
0.9 \gev and 1.6 \gev. 

Moments of inclusive distributions and rates for semileptonic and rare $B$ decays can be related via Operator Product Expansions (OPE) \cite{Chay:1990da} 
to fundamental parameters of the Standard
Model, such as the Cabibbo-Kobayashi-Maskawa matrix elements \Vcb and \Vub~\cite{ckm} and the heavy quark masses $m_b$ and $m_c$. 
These expansions in $1/m_b$ 
and the strong coupling constant $\alpha_s$ 
involve non-perturbative quantities that can be extracted from  moments of inclusive distributions. We plan to use measurements of the hadron mass and lepton energy moments~\cite{lepton}
to improve the determination of \Vcb from the semileptonic decay rate~\cite{interpret}.

\ifnum\isprl<1
\section{The \babar\ Dataset and Monte Carlo Simulation}
\label{sec:data}
\fi
\ifnum\isprl<1
\subsection{Data}
\fi

The measurement presented here is based on a sample of 89 million \BB\ pairs
collected  on the \FourS\ resonance by the \babar\ detector~\cite{Aubert:2001tu} at the
PEP-II  asymmetric-energy  $e^+e^-$ storage ring operating at SLAC.
\ifnum\isprl<1
\subsection{Monte Carlo Simulation}
\label{subsec:montecarlo}
\fi
We use Monte Carlo (MC) simulations of the \babar\ detector based on
\geantf~\cite{Agostinelli:2002hh} to determine background distributions
and to correct for detector acceptance effects. The simulations of 
$\Bxclnu$ decays use a parameterization of form factors for $\Bb\to
D^{*}\ell^-\nub$~\cite{Duboscq:1996mv}, and models for   
$\Bb\to D \ell^-\nub,D^{**}\ell^-\nub$~\cite{ISGW2} and 
$\Bb\to D \pi \ell^-\nub, D^* \pi \ell^-\nub$~\cite{J.GoityandW.Roberts}. 
\ifnum\isprl<1
\section{Data Analysis}
\label{sec:analysis}
\fi

\ifnum\isprl<1
\subsection{Selection of Hadronic $B$ Decays, $B \ra \Db Y$}
\fi

The analysis uses \FourS\to\BB\ events in which one of the \B mesons decays to hadrons and is fully reconstructed ($B_{\rm reco}$) and the semileptonic 
decay of the recoiling \Bb\ meson ($B_{\rm recoil}$) is identified by the presence of an electron or muon. While this approach 
results in a low overall event selection efficiency, it allows for the determination of the momentum, charge, and flavor of the \B mesons. 
To obtain a large sample of $B$ mesons,  many exclusive hadronic decays are reconstructed~\cite{Aubert:2003zw}.  
The kinematic consistency of these $B_{\rm reco}$ candidates 
is checked with two variables,
the beam-energy-substituted mass $\mes = \sqrt{s/4 -
\vec{p}^{\,2}_B}$ and the energy difference 
$\Delta E = E_B - \sqrt{s}/2$. Here $\sqrt{s}$ is the total
energy in the center of mass frame (c.m.), $\vec{p}_B$ and $E_B$
denote the c.m. momentum and c.m. energy of the $B_{\rm reco}$ candidate.  
We require $\Delta E = 0$ within three standard
deviations as measured for each mode.
For a given \breco\ decay mode, the purity 
is estimated as the signal fraction in events  with \mes$>5.27$\gev.
For events with one high-momentum lepton the purity is approximately 70\%. 

\ifnum\isprl<1
\subsection{Selection of Semileptonic Decays, $\Bxlnu$}
\fi

Semileptonic  decays 
are identified by the presence of one and only one electron or muon above a minimum momentum $p^*_{\rm min}$  measured in the rest frame of the $B_{\rm recoil}$ meson recoiling against the $B_{\rm reco}$. 
Electrons are selected~\cite{Aubert:2002uf} with 92\% average efficiency and a
hadron misidentification rate ranging between 0.05\% and 0.1\%.
Muons are identified~\cite{Aubert:2001tu} with an
efficiency ranging between 60\% ($p_{\rm lab}=1\gev$) and 75\% ($p_{\rm lab}>2\gev$)
and a hadron misidentification rate between 1\% and 3\%.
Efficiencies and misidentification rates are estimated 
from selected samples of electrons, muons, pions, and kaons.
We impose the condition $Q_b Q_{\ell} < 0$, where $Q_{\ell}$ is the charge of the lepton and $Q_b$ is the 
charge of the  $b$-quark of the $B_{\rm reco}$.         
This condition is fulfilled for primary leptons, except for \BzBzb\ events in which flavor mixing has occurred.
We require the total observed charge of the event to be $|Q_{\rm tot}|= |Q_{\rm B_{reco}} + Q_{\rm B_{recoil}}| \leq 1$, 
allowing for a charge imbalance in events with low momentum tracks or photon conversions.

The hadronic system $X$ in the decay \Bxlnu\ is reconstructed from charged
tracks  and energy depositions
in the calorimeter that are not associated with
the \breco\ candidate or the charged lepton. Depending on particle identification information the charged tracks are assigned either the $K^{\pm}$ or $\pi^{\pm}$ mass.
Procedures are implemented to eliminate 
fake charged tracks, low-energy beam-generated photons, and energy depositions in the calorimeter from charged and neutral hadrons.  

The neutrino four-momentum $p_{\nu}$ is estimated from the
missing four-momentum  $p_{\rm miss} = p_{\FourS}-p_{\breco} -p_X-p_\ell$, 
where all momenta are measured in the laboratory frame. 
The measured $p_{\rm miss}$
is an important indicator of the quality of the reconstruction of $X$.
We impose the following criteria:  $E_{\rm miss} > 0.5 \gev$, $|\vec {p}_{\rm miss}| > 0.5 \gev$, and $|E_{\rm miss} - |\vec {p}_{\rm miss}|| < 0.5 \gev$.
The mass of the hadronic system $M_X$ is determined by a
kinematic fit that imposes four-momentum conservation, the equality of
the masses of the two $B$ mesons, and constrains $p_{\nu}^2 = 0$.  The resulting mean resolution in $M_X$ is 350 \mev.
 \begin{figure}[t]
    \begin{centering}
	\epsfig{file=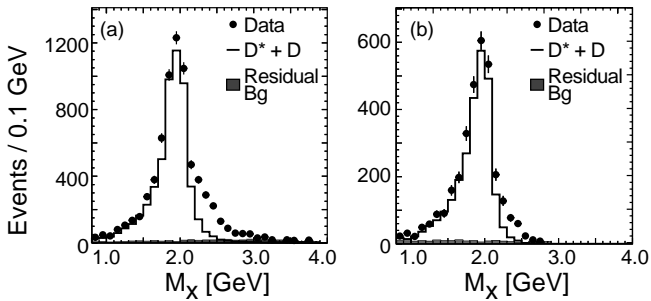,height=4.0cm}
    \caption{$M_X$ distributions after subtraction of the \breco\ background, for (a) $\psm = 0.9~\gev$, and (b) $\psm = 1.6~\gev$. The Monte Carlo prediction for decays to $D$ and $D^*$ is indicated by the open histogram, the small residual background by the solid histogram.
    \label{fig:mx}}
   \end{centering}
\vspace{-0.3cm}
   \end{figure} 

The background is dominated by combinatorial background in the \breco\ sample. To estimate this
background we fit the observed $\mes$  distribution to a sum of an empirical function~\cite{Albrecht:1987nr} 
describing the combinatorial background from both continuum and \BB\ events and a narrow signal function~\cite{Skwarnicki:1986xj}
peaked at the $B$ meson mass. This fit is performed separately for several bins in $M_X$, thus
accounting for changes in background as a function of $M_X$. For $p^*_{\rm min}=0.9 \gev$ and $\mes > 5.27 \gev$, 
we find a total of $7114$ signal events above a 
combinatorial
background of $2102$ events. Figure~\ref{fig:mx} shows $M_X$ distributions after  
\breco\ background subtraction. 
The dominant contributions are from the lowest mass mesons, 
($D^+$, $D^0$) and ($D^{*+}$, $D^{*0}$), but there are clear indications
for higher mass states.

The residual background, estimated from MC simulation, is due to hadron misidentification,
$\tau^{\pm}$ leptons, $\Bxulnu$ decays, and secondary leptons from semileptonic decays of 
$D^{(*)}$ and $D_s$ mesons, either from 
\BzBzb\ mixed events or produced in $b \ra c \cbar s$ transitions.

 \begin{figure}[t]
    \begin{centering}
	\epsfig{file=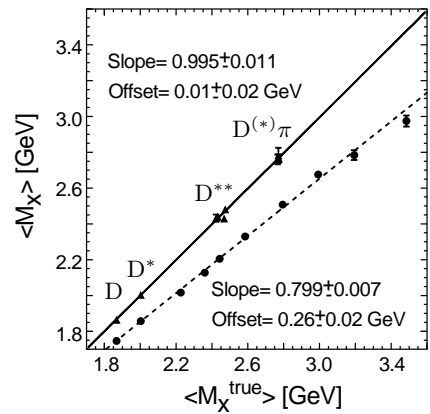,height=5.5cm}
\unitlength1.0cm
\put(-4.3,1.6){$\rm D$}
\put(-4.0,1.9){$\rm D^*$}
\put(-3.1,2.9){$\rm D^{**}$}
\put(-2.6,3.75){$\rm D^{(*)}\pi$}
     \caption{ \label{fig:calib} Results of the $\langle M_X \rangle$ calibration procedure.
The calibration data and fit results are shown by the lower dashed line (circles), the verification by the upper solid line (triangles).   }
   \end{centering}
\vspace{-0.3cm}
   \end{figure}

To extract 
unbiased moments \mmxn, 
we need to correct for effects that can distort the mass distributions (see also~\cite{Bauer:2002va}).
We use observed linear relationships between the measured \mmxn\ and generated $\langle M_X^{n~{\rm true}} \rangle$ values from MC simulations in bins of $M_X^{n~{\rm true}}$ (see Fig.~\ref{fig:calib}) to calibrate the measurement of 
$M_X^n$ on an event-by-event basis.
Since any radiative photon is included in the measured hadron mass and our definition of $M_X$ does not include these photons, we employ PHOTOS~\cite{Barberio:1994qi}
to simulate QED radiative effects and correct for their impact (less than 5\%) on the moments as part of the calibration procedure. 

To verify this procedure, we apply the calibration to the 
measured masses for individual hadronic states in simulated $\Bxclnu$  decays, 
and compare their calibrated mass moments to the true mass moments. 
The result of this test is also shown in Fig.~\ref{fig:calib} for $M_X$, indicating that the calibration reproduces the true moments over the full mass range.  
Similar curves are obtained for $M_X^2$, $M_X^3$, and $M_X^4$.  
We observe 
no significant mass bias after calibration.  
The MC-based calibration procedure has also been validated on a data sample of partially reconstructed $D^*$ decays.  

Detailed studies show that the slope and offset of the calibration curves vary slightly as a function of the multiplicity of the hadron system and as a function of $ E_{\rm miss} -|\vec {p}_{\rm miss}|$. Thus, instead of one universal calibration curve for all data, we split the data into three bins in multiplicity and three bins in $E_{\rm miss} -|\vec {p}_{\rm miss}|$, and derive a total of nine calibration curves, one for each subsample.
\begin{figure*}[t]
    \begin{centering}
	\epsfig{file=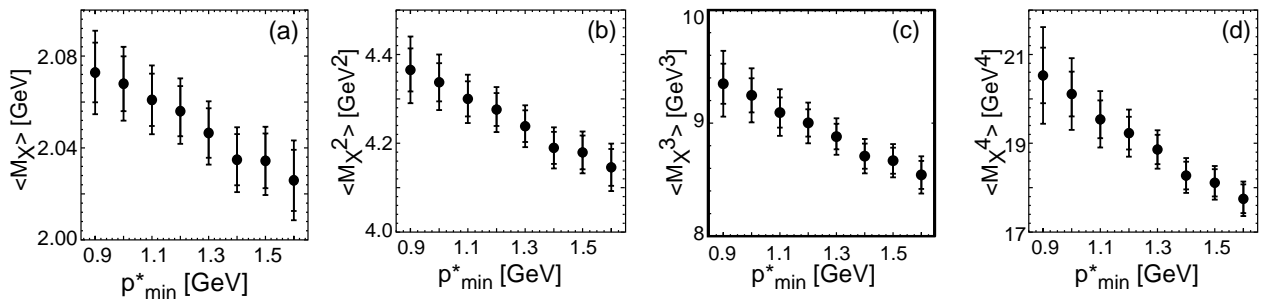,height=4.0cm} 
   \end{centering}
    \caption{
    \label{fig:results} Measured moments (a) \mmx, (b) \mmxs, (c) \mmxc, and (d) \mmxq\  for different lepton momenta, $p^*_{\rm min}$. The bars indicate the statistical and total errors. The individual moments are highly correlated.}
   \end{figure*}
We estimate and subtract the contribution to the moments 
from residual backgrounds and then correct the result by a factor ${\cal C}_n$ 
for the effect of detection and selection efficiencies.
We can express the fully corrected hadronic mass moment $\mmxn$ as
\begin{equation}
\label{equ:mx}
    \mmxn = \frac{\mmxn^{\rm DATA}_{\rm calib} - f_{\rm bg} \cdot \mmxn^{\rm MCbg}_{\rm calib}} 
{1-f_{\rm bg}} \times {\cal C}_n  ,
\end{equation}
where $\mmxn^{\rm DATA}_{\rm calib}$ and $\mmxn^{\rm MCbg}_{\rm calib}$ are the calibrated moments of the data 
and the residual background. 
The factor $f_{\rm bg}$ denotes the size of the residual background relative 
to the data.
  
Decays to higher mass final states usually generate higher multiplicities and are more strongly affected by the requirements on $E_{\rm miss}$ and $|\vec{p}_{\rm miss}|$
due to limited efficiency.
In addition, the different decay modes have different spin configurations and thus different angular distributions. 
The correction factor ${\cal C}_n$ in Eq.~\ref{equ:mx} accounts for these effects.
It is determined by MC simulation, and is found to be within 1\% of unity. 

The hadronic mass moments 
\mmxn\
obtained after background subtraction, correction for $\Bxulnu$ decays, and mass calibration are presented in Fig.~\ref{fig:results} as a function of $p^*_{\rm min}$.  
The measurements are highly correlated. 
The numerical results and the full correlation matrix for the four sets of $p^*_{\rm min}$ dependent moment measurements can be found in Tables II-VIII.
The four moments increase as $p^*_{\rm min}$ decreases due
to the presence of higher mass charm states.
Fits to the $p^*_{\rm min}$ dependence assuming constant moments are inconsistent with our results, 
with $\chi^2$ probabilities less than~$0.4\%$. 
\ifnum\isprl<1
\section{Systematic}
\label{sec:sus}
\fi

Table \ref{tab:results} shows the four measured moments and their principal errors for 
$p^*_{\rm min}= 0.9~\gev$ and $p^*_{\rm min}= 1.6~\gev$.
The main sources of systematic errors are the precision in the modeling of 
the detector efficiency and particle reconstruction, the subtraction of the combinatorial background 
of the $B_{\rm reco}$ sample, the residual background estimate, and the uncertainties  in the modeling of the hadronic states.
The uncertainty related to the detector modeling and event reconstruction 
has been estimated by MC
simulations of the track and photon efficiencies. Resolutions, fake rates, and background rates have 
been studied in detail by varying the adjustments to the MC simulation that are introduced to improve the agreement with data. 
The uncertainty in the combinatorial $B_{\rm reco}$ background subtraction is estimated by varying the lower limit 
of the signal region in the $m_{\rm ES}$ distribution. 
The error due to the subtraction of the residual background is dominated by the
uncertainties 
(typically 30\% \cite{PDG2002}) 
in the rate of $D^{(*)}$ and $D_s$ production via $b \ra c \cbar s$ transitions. 
The uncertainty related to the modeling of the semileptonic $B$ decays
is estimated by varying the branching fractions, 
in particular those for the high mass resonant and non-resonant states. 
Uncertainties in the radiative corrections, especially effects not included in PHOTOS, 
are estimated by removing photons of above a variable energy limit from the hadronic system $X$. 

To test the stability of the moment measurements,
the data are divided into several independent subsamples: $B^{\pm}$ and $B^0$, decays to electrons and muons,
different run periods, positive and negative $E_{\rm miss} - |\vec{p}_{\rm miss}|$, and 
              high and low purity \breco\ modes. No significant variations are observed.
  
In summary, we have performed a measurement of the 
first four moments \mmxn\
of the hadronic mass distribution in semileptonic $B$ decays.
For $p^*_{\rm min}=1.5~\gev$, our measurement of $\mmxs = 4.18 \pm 0.04 (stat.) \pm 0.03 (syst.) \gev^2$ 
agrees well with the single result from CLEO \cite{Cronin-Hennessy:2001fk}. 
The selection of events with one fully reconstructed hadronic $B$ decay, the kinematic fit, and calibration of the hadronic mass in the semileptonic decay of the second $B$ decay 
have led to moment measurements with comparable 
statistical and systematic errors. The results do not depend on assumptions for branching fractions and mass
distributions for higher mass hadronic states. 
The measured  moments increase significantly as the limit on the lepton momentum, $p^*_{\rm min}$, is lowered, as expected for
increasing contributions from higher mass states.
The set of moments presented here can be used to test the applicability of the OPE to semileptonic and rare $B$ decays.
Combining them with the measured semileptonic decay rate is expected to result in a significantly improved
determination of \Vcb \cite{interpret,lepton}.

\begin{table*}[htbp]
\caption[]{\label{tab:results} {Results for \mmxn\ for the two extreme values of $p^*_{\rm min}$, with statistical and systematic errors and details on the major contributions to the systematic uncertainties. 
\\}}
\begin{tabular}{c c c c c c c c c c}
\hline\hline
&                                                   & \multicolumn{3}{c}{}                                         & Detector  & \breco     & Residual  &\Bxclnu    & Radiative \\
& \raisebox{1.7ex}[-1.7ex]{$p^*_{\rm min}$ (\gev)}  & \multicolumn{3}{c}{\raisebox{1.7ex}[-1.7ex]{\mmxn (\gevn)}}  & Response & Background & Background &Model      & Corrections \\ \hline  
$n=1$ & 0.9 &  2.073 & $\pm$ 0.013 & $\pm$ 0.013  & 0.009  & 0.004 & 0.008 & 0.002 & 0.003  \\ 
      & 1.6 &  2.026 & $\pm$ 0.013 & $\pm$ 0.012  & 0.010  & 0.004 & 0.002 & 0.002 & 0.004  \\ \hline
$n=2$ & 0.9 &  4.366 & $\pm$ 0.049 & $\pm$ 0.058  & 0.034  & 0.023 & 0.039 & 0.009 & 0.009  \\ 
      & 1.6 &  4.146 & $\pm$ 0.042 & $\pm$ 0.036  & 0.031  & 0.009 & 0.007 & 0.007 & 0.013  \\ \hline
$n=3$ & 0.9 &  9.35  & $\pm$ 0.18  & $\pm$ 0.23   & 0.15   & 0.05  & 0.16  & 0.01  & 0.03   \\ 
      & 1.6 &  8.54  & $\pm$ 0.12  & $\pm$ 0.11   & 0.10   & 0.02  & 0.01  & 0.01  & 0.04   \\ \hline
$n=4$ & 0.9 & 20.53  & $\pm$ 0.63  & $\pm$ 0.90   & 0.58   & 0.31  & 0.58  & 0.13  & 0.14   \\ 
      & 1.6 & 17.75  & $\pm$ 0.32  & $\pm$ 0.23   & 0.19   & 0.06  & 0.02  & 0.08  & 0.09   \\ 
 \hline\hline
\end{tabular}
\end{table*}

We are grateful for the excellent luminosity and machine conditions
provided by our \pep2\ colleagues, 
and for the substantial dedicated effort from
the computing organizations that support \babar.
The collaborating institutions wish to thank 
SLAC for its support and kind hospitality. 
This work is supported by
DOE
and NSF (USA),
NSERC (Canada),
IHEP (China),
CEA and
CNRS-IN2P3
(France),
BMBF and DFG
(Germany),
INFN (Italy),
FOM (The Netherlands),
NFR (Norway),
MIST (Russia), and
PPARC (United Kingdom). 
Individuals have received support from the 
A.~P.~Sloan Foundation, 
Research Corporation,
and Alexander von Humboldt Foundation.


\begin{thebibliography}{24}
\bibitem[{cha()}]{chargecon}
\bibinfo{note}{Charge conjugation is implied throughout this Letter.}
\end{thebibliography}

\begin{table*}[htbp]
\caption[]{\label{tab:mom1} {\small Results for \mmx\ for different values of $p^*_{\rm min}$, with statistical and total systematic errors. The last five columns show separately the five dominant contributions to the systematic uncertainty: detector response, combinatorial $B_{\rm reco}$ background, residual background subtraction, dependence on the $\Bxclnu$ decay model, and radiative corrections.\\}}
\begin{tabular}{c c c c c c c c c}
\hline\hline
            &\multicolumn{3}{c}{\mmx ~(\gev)}            & Detector &  \breco    & Residual   & \Bxclnu & Radiative\\ 
\raisebox{1.7ex}[-1.7ex]{\psm (\gev)} & & stat. &sys.      & Response & Background & Background & Model   & Corrections\\ \hline
 0.9 &  2.073& $\pm$ 0.013& $\pm$ 0.013 & 0.009 & 0.004 & 0.008& 0.002&0.003 \\ 
 1.0 &  2.068& $\pm$ 0.012& $\pm$ 0.012 & 0.009 & 0.003 & 0.005& 0.002&0.004\\ 
 1.1 &  2.061& $\pm$ 0.011& $\pm$ 0.011 & 0.009 & 0.002 & 0.003& 0.002&0.005\\ 
 1.2 &  2.056& $\pm$ 0.011& $\pm$ 0.010 & 0.008 & 0.001 & 0.002& 0.003&0.004\\ 
 1.3 &  2.047& $\pm$ 0.011& $\pm$ 0.010 & 0.008 & 0.001 & 0.002& 0.002&0.004\\ 
 1.4 &  2.035& $\pm$ 0.011& $\pm$ 0.010 & 0.008 & 0.002 & 0.002& 0.002&0.005\\ 
 1.5 &  2.034& $\pm$ 0.012& $\pm$ 0.010 & 0.008 & 0.003 & 0.001& 0.002&0.004\\ 
 1.6 &  2.026& $\pm$ 0.013& $\pm$ 0.012 & 0.010 & 0.004 & 0.002& 0.002&0.004\\ \hline\hline
\end{tabular}
\end{table*}

\begin{table*}[ht]
\begin{center}
\caption[]{\label{tab:mom2} {\small Results for \mmxs\ for different values of $p^*_{\rm min}$, with statistical and total systematic errors. The last five columns show separately the five dominant contributions to the systematic uncertainty: detector response, combinatorial $B_{\rm reco}$ background, residual background subtraction, dependence on the $\Bxclnu$ decay model, and radiative corrections.\\}}
\begin{tabular}{c c c c c c c c c}
\hline\hline
            &\multicolumn{3}{c}{\mmxs ~($\gev^2$)}  & Detector & \breco     & Residual   & \Bxclnu & Radiative\\ 
\raisebox{1.7ex}[-1.7ex]{\psm (\gev)} & & stat. &sys. & Response & Background & Background & Model   & Corrections\\ \hline  
 0.9 &  4.366& $\pm$ 0.049& $\pm$ 0.058 & 0.034  & 0.023 & 0.039& 0.009&0.009 \\ 
 1.0 &  4.338& $\pm$ 0.043& $\pm$ 0.048 & 0.033  & 0.016 & 0.025& 0.009&0.015 \\ 
 1.1 &  4.300& $\pm$ 0.040& $\pm$ 0.042 & 0.032  & 0.006 & 0.016& 0.010&0.019 \\ 
 1.2 &  4.276& $\pm$ 0.037& $\pm$ 0.039 & 0.030  & 0.006 & 0.011& 0.011&0.017 \\ 
 1.3 &  4.239& $\pm$ 0.036& $\pm$ 0.035 & 0.028  & 0.006 & 0.007& 0.010&0.016 \\ 
 1.4 &  4.190& $\pm$ 0.036& $\pm$ 0.035 & 0.027  & 0.007 & 0.005& 0.008&0.019 \\ 
 1.5 &  4.180& $\pm$ 0.038& $\pm$ 0.031 & 0.026  & 0.006 & 0.005& 0.008&0.014 \\ 
 1.6 &  4.146& $\pm$ 0.042& $\pm$ 0.036 & 0.031  & 0.009 & 0.007& 0.007&0.013 \\ \hline\hline
\end{tabular}
\end{center}
\end{table*}

\begin{table*}[ht]
\begin{center}
\caption[]{\label{tab:mom3} {\small Results for \mmxc\ for different values of $p^*_{\rm min}$, with statistical and total systematic errors. The last five columns show separately the five dominant contributions to the systematic uncertainty: detector response, combinatorial $B_{\rm reco}$ background, residual background subtraction, dependence on the $\Bxclnu$ decay model, and radiative corrections.\\}}
\begin{tabular}{c c c c c c c c c}
\hline\hline
            &\multicolumn{3}{c}{\mmxc ~($\gev^3$)}            & Detector & \breco   & Residual     & \Bxclnu & Radiative\\ 
\raisebox{1.7ex}[-1.7ex]{\psm (\gev)} & & stat. &sys.        & Response & Background & Background & Model   & Corrections\\ \hline
 0.9 &  9.35& $\pm$ 0.18 $\pm$ &0.23 & 0.15  &0.05 & 0.16&0.01&0.03\\ 
 1.0 &  9.25& $\pm$ 0.15 $\pm$ &0.19 & 0.14  &0.05 & 0.10&0.01&0.04\\ 
 1.1 &  9.09& $\pm$ 0.13 $\pm$ &0.16 & 0.13  &0.03 & 0.07&0.01&0.06\\ 
 1.2 &  9.00& $\pm$ 0.12 $\pm$ &0.13 & 0.11  &0.02 & 0.05&0.01&0.05\\ 
 1.3 &  8.88& $\pm$ 0.11 $\pm$ &0.12 & 0.10  &0.01 & 0.02&0.01&0.05\\ 
 1.4 &  8.71& $\pm$ 0.11 $\pm$ &0.10 & 0.09  &0.02 & 0.01&0.01&0.05\\ 
 1.5 &  8.67& $\pm$ 0.12 $\pm$ &0.09 & 0.08  &0.02 & 0.01&0.01&0.04\\ 
 1.6 &  8.54& $\pm$ 0.12 $\pm$ &0.11 & 0.10  &0.02 & 0.01&0.01&0.04\\ \hline\hline
\end{tabular}
\end{center}
\end{table*}

\begin{table*}[ht]
\begin{center}
\caption[]{\label{tab:mom4} {\small Results for \mmxq\ for different values of $p^*_{\rm min}$, with statistical and total systematic errors. The last five columns show separately the five dominant contributions to the systematic uncertainty: detector response, combinatorial $B_{\rm reco}$ background, residual background subtraction, dependence on the $\Bxclnu$ decay model, and radiative corrections.\\}}
\begin{tabular}{c c c c c c c c c}
\hline\hline
            &\multicolumn{3}{c}{\mmxq ~($\gev^4$)}         & Detector & \breco     & Residual   & \Bxclnu & Radiative\\ 
\raisebox{1.7ex}[-1.7ex]{\psm (\gev)} & & stat. &sys.        & Response & Background & Background & Model   & Corrections\\ \hline
 0.9 & 20.53& $\pm$ 0.63& $\pm$ 0.90& 0.58& 0.31& 0.58& 0.13&0.14\\ 
 1.0 & 20.11& $\pm$ 0.51& $\pm$ 0.64& 0.47& 0.20& 0.36& 0.11&0.11\\ 
 1.1 & 19.54& $\pm$ 0.43& $\pm$ 0.50& 0.38& 0.08& 0.23& 0.13&0.18\\ 
 1.2 & 19.23& $\pm$ 0.37& $\pm$ 0.41& 0.32& 0.07& 0.15& 0.13&0.15\\ 
 1.3 & 18.86& $\pm$ 0.33& $\pm$ 0.31& 0.23& 0.04& 0.08& 0.12&0.13\\ 
 1.4 & 18.28& $\pm$ 0.31& $\pm$ 0.28& 0.21& 0.04& 0.04& 0.10&0.15\\ 
 1.5 & 18.11& $\pm$ 0.31& $\pm$ 0.24& 0.19& 0.05& 0.02& 0.09&0.09\\ 
 1.6 & 17.75& $\pm$ 0.32& $\pm$ 0.23& 0.19& 0.06& 0.02& 0.08&0.09\\ \hline\hline
\end{tabular}
\end{center}
\end{table*}

\begin{table*}
\begin{center}
{\small
\caption[]{\label{correl1}{\small Correlation Coefficients for \mmx\ and \mmxs\ measurements with different \psm cuts. (Tables with higher precision can be obained from the authors.)\\}}
\begin{tabular}{cc|cccccccc|cccccccc}
\hline\hline
\multicolumn{2}{c|}{\vvsmallrule \psm} & \multicolumn{8}{c|}{\mmx}& \multicolumn{8}{c}{\mmxs}\\ 
\multicolumn{2}{c|}{(\gev)} & 0.9 &  1.0 &  1.1 &  1.2 &  1.3 &  1.4 &  1.5 &  1.6 & 0.9 &  1.0 &  1.1 &  1.2 &  1.3 &  1.4 &  1.5 & 1.6\\ \hline 
& 0.9 & 1.00 & 0.91 & 0.85 & 0.79 & 0.74 & 0.70 & 0.67 & 0.63 & 0.96 & 0.88 & 0.82 & 0.77 & 0.72 & 0.68 & 0.64 & 0.61\\
& 1.0 &      & 1.00 & 0.93 & 0.87 & 0.81 & 0.77 & 0.73 & 0.69 & 0.84 & 0.96 & 0.89 & 0.84 & 0.78 & 0.74 & 0.70 & 0.66\\
& 1.1 &      &      & 1.00 & 0.93 & 0.87 & 0.82 & 0.78 & 0.75 & 0.74 & 0.85 & 0.96 & 0.90 & 0.84 & 0.80 & 0.76 & 0.72\\ 
& 1.2 &      &      &      & 1.00 & 0.94 & 0.88 & 0.84 & 0.80 & 0.66 & 0.77 & 0.87 & 0.97 & 0.90 & 0.85 & 0.81 & 0.77\\
\raisebox{1.7ex}[-1.7ex]{\mmx}& 1.3 &      &      &      &      & 1.00 & 0.94 & 0.90 & 0.85 & 0.60 & 0.69 & 0.78 & 0.87 & 0.97 & 0.91 & 0.87 & 0.82\\ 
& 1.4 &      &      &      &      &      & 1.00 & 0.95 & 0.90 & 0.55 & 0.63 & 0.71 & 0.79 & 0.88 & 0.97 & 0.92 & 0.87\\
& 1.5 &      &      &      &      &      &      & 1.00 & 0.95 & 0.51 & 0.58 & 0.66 & 0.73 & 0.81 & 0.89 & 0.96 & 0.91\\
& 1.6 &      &      &      &      &      &      &      & 1.00 & 0.46 & 0.54 & 0.61 & 0.68 & 0.75 & 0.82 & 0.89 & 0.96\\ \hline 
& 0.9 &      &      &      &      &      &      &      &      & 1.00 & 0.87 & 0.77 & 0.69 & 0.62 & 0.57 & 0.52 & 0.48\\ 
& 1.0 &      &      &      &      &      &      &      &      &      & 1.00 & 0.88 & 0.79 & 0.72 & 0.65 & 0.60 & 0.56\\
& 1.1 &      &      &      &      &      &      &      &      &      &      & 1.00 & 0.90 & 0.81 & 0.74 & 0.68 & 0.63\\ 
& 1.2 &      &      &      &      &      &      &      &      &      &      &      & 1.00 & 0.90 & 0.82 & 0.76 & 0.70\\
\raisebox{1.7ex}[-1.7ex]{\mmxs}& 1.3 &      &      &      &      &      &      &      &      &      &      &      &      & 1.00 & 0.91 & 0.84 & 0.78\\
& 1.4 &      &      &      &      &      &      &      &      &      &      &      &      &      & 1.00 & 0.93 & 0.86\\ 
& 1.5 &      &      &      &      &      &      &      &      &      &      &      &      &      &      & 1.00 & 0.92\\
& 1.6 &      &      &      &      &      &      &      &      &      &      &      &      &      &      &      & 1.00\\ \hline \hline
\end{tabular}
}
\end{center}
\end{table*}

\begin{table*}
\begin{center}
{\small
\caption[]{\label{correl2}{\small Correlation Coefficients for \mmx, \mmxs, \mmxc, and \mmxq\ measurements with different \psm cuts. (Tables with higher precision can be obained from the authors.)\\}}
\begin{tabular}{cc|cccccccc|cccccccc}
\hline\hline
\multicolumn{2}{c|}{\vvsmallrule \psm} & \multicolumn{8}{c|}{\mmxc}& \multicolumn{8}{c}{\mmxq}\\ 
\multicolumn{2}{c|}{(\gev)} & 0.9 &  1.0 &  1.1 &  1.2 &  1.3 &  1.4 &  1.5 &  1.6 & 0.9 &  1.0 &  1.1 &  1.2 &  1.3 &  1.4 &  1.5 & 1.6\\ \hline
& 0.9 & 0.89 & 0.81 & 0.76 & 0.71 & 0.67 & 0.63 & 0.60 & 0.57 & 0.80 & 0.73 & 0.68 & 0.64 & 0.61 & 0.58 & 0.55 & 0.52\\
& 1.0 & 0.73 & 0.89 & 0.83 & 0.78 & 0.73 & 0.69 & 0.66 & 0.62 & 0.62 & 0.80 & 0.75 & 0.70 & 0.66 & 0.63 & 0.60 & 0.57\\
& 1.1 & 0.61 & 0.75 & 0.89 & 0.84 & 0.79 & 0.74 & 0.71 & 0.67 & 0.50 & 0.64 & 0.80 & 0.76 & 0.71 & 0.68 & 0.65 & 0.62\\
& 1.2 & 0.53 & 0.65 & 0.77 & 0.90 & 0.84 & 0.80 & 0.76 & 0.72 & 0.41 & 0.53 & 0.66 & 0.81 & 0.76 & 0.73 & 0.70 & 0.66\\
\raisebox{1.7ex}[-1.7ex]{\mmx}& 1.3 &0.46 & 0.56 & 0.67 & 0.78 & 0.90 & 0.85 & 0.81 & 0.77 & 0.35 & 0.44 & 0.56 & 0.69 & 0.82 & 0.78 & 0.74 & 0.71\\
& 1.4 & 0.41 & 0.50 & 0.59 & 0.69 & 0.79 & 0.90 & 0.86 & 0.81 & 0.29 & 0.38 & 0.48 & 0.58 & 0.70 & 0.83 & 0.79 & 0.75\\
& 1.5 & 0.37 & 0.45 & 0.54 & 0.63 & 0.72 & 0.82 & 0.90 & 0.85 & 0.26 & 0.34 & 0.43 & 0.52 & 0.62 & 0.74 & 0.83 & 0.79\\  
& 1.6 & 0.34 & 0.41 & 0.49 & 0.57 & 0.65 & 0.74 & 0.82 & 0.90 & 0.23 & 0.30 & 0.38 & 0.47 & 0.55 & 0.66 & 0.74 & 0.83\\ \hline
& 0.9 & 0.98 & 0.85 & 0.75 & 0.67 & 0.61 & 0.56 & 0.51 & 0.48 & 0.93 & 0.80 & 0.71 & 0.64 & 0.58 & 0.53 & 0.49 & 0.45\\
& 1.0 & 0.80 & 0.98 & 0.86 & 0.78 & 0.70 & 0.64 & 0.59 & 0.55 & 0.72 & 0.92 & 0.81 & 0.74 & 0.67 & 0.61 & 0.57 & 0.52\\
& 1.1 & 0.67 & 0.82 & 0.98 & 0.88 & 0.80 & 0.73 & 0.67 & 0.62 & 0.57 & 0.73 & 0.92 & 0.83 & 0.75 & 0.69 & 0.64 & 0.59\\
& 1.2 & 0.58 & 0.70 & 0.84 & 0.98 & 0.89 & 0.81 & 0.75 & 0.69 & 0.47 & 0.60 & 0.76 & 0.93 & 0.84 & 0.77 & 0.71 & 0.66\\
\raisebox{1.7ex}[-1.7ex]{\mmxs}& 1.3 & 0.50 & 0.61 & 0.73 & 0.85 & 0.98 & 0.89 & 0.83 & 0.76 & 0.39 & 0.50 & 0.64 & 0.78 & 0.93 & 0.85 & 0.79 & 0.73\\
& 1.4 & 0.44 & 0.54 & 0.65 & 0.75 & 0.86 & 0.98 & 0.91 & 0.84 & 0.33 & 0.43 & 0.54 & 0.66 & 0.79 & 0.94 & 0.87 & 0.80\\
& 1.5 & 0.40 & 0.49 & 0.59 & 0.68 & 0.78 & 0.89 & 0.98 & 0.91 & 0.30 & 0.38 & 0.48 & 0.59 & 0.70 & 0.83 & 0.94 & 0.87\\
& 1.6 & 0.37 & 0.45 & 0.53 & 0.62 & 0.71 & 0.81 & 0.89 & 0.98 & 0.27 & 0.34 & 0.43 & 0.53 & 0.63 & 0.75 & 0.84 & 0.94\\ \hline\hline
\end{tabular}
}
\end{center}
\end{table*}

\begin{table*}
\begin{center}
{\small
\caption[]{\label{correl3}{\small Correlation Coefficients for \mmxc\ and \mmxq\ measurements with different \psm cuts. (Tables with higher precision can be obained from the authors.)\\}}
\begin{tabular}{cc|cccccccc|cccccccc}
\hline\hline
\multicolumn{2}{c|}{\vvsmallrule \psm} & \multicolumn{8}{c|}{\mmxc}& \multicolumn{8}{c}{\mmxq}\\ 
\multicolumn{2}{c|}{(\gev)} & 0.9 &  1.0 &  1.1 &  1.2 &  1.3 &  1.4 &  1.5 &  1.6 & 0.9 &  1.0 &  1.1 &  1.2 &  1.3 &  1.4 &  1.5 & 1.6\\ \hline 
& 0.9 & 1.00 & 0.82 & 0.69 & 0.59 & 0.52 & 0.45 & 0.41 & 0.37 & 0.98 & 0.81 & 0.68 & 0.58 & 0.51 & 0.45 & 0.41 & 0.37\\
& 1.0 &      & 1.00 & 0.84 & 0.72 & 0.63 & 0.55 & 0.50 & 0.46 & 0.77 & 0.98 & 0.82 & 0.71 & 0.62 & 0.54 & 0.49 & 0.45\\
& 1.1 &      &      & 1.00 & 0.86 & 0.75 & 0.66 & 0.60 & 0.54 & 0.61 & 0.78 & 0.98 & 0.85 & 0.74 & 0.65 & 0.59 & 0.54\\
& 1.2 &      &       &     & 1.00 & 0.87 & 0.77 & 0.70 & 0.63 & 0.50 & 0.64 & 0.80 & 0.98 & 0.86 & 0.75 & 0.69 & 0.62\\
\raisebox{1.7ex}[-1.7ex]{\mmxc}& 1.3 &    &  &  &   &            1.00 & 0.88 & 0.80 & 0.73 & 0.42 & 0.53 & 0.67 & 0.83 & 0.98 & 0.87 & 0.79 & 0.72\\
& 1.4 &      &       &     &       &     & 1.00 & 0.91 & 0.83 & 0.35 & 0.45 & 0.57 & 0.70 & 0.83 & 0.99 & 0.90 & 0.82\\
& 1.5 &      &        &     &      &     &     &  1.00 & 0.91 & 0.31 & 0.40 & 0.51 & 0.62 & 0.74 & 0.88 & 0.99 & 0.90\\
& 1.6 &     &        &     &       &     &     &      &  1.00 & 0.28 & 0.36 & 0.45 & 0.56 & 0.66 & 0.79 & 0.89 & 0.99\\ \hline
& 0.9 &     &        &     &       &     &     &      &      &  1.00 & 0.78 & 0.62 & 0.50 & 0.42 & 0.36 & 0.32 & 0.28\\
& 1.0 &     &        &     &       &     &     &      &      &      &  1.00 & 0.79 & 0.65 & 0.54 & 0.46 & 0.41 & 0.36\\
& 1.1 &     &        &     &       &     &     &      &      &       &     &  1.00 & 0.82 & 0.69 & 0.58 & 0.51 & 0.46\\
& 1.2 &     &        &     &       &     &     &      &      &       &     &      &  1.00 & 0.84 & 0.71 & 0.63 & 0.56\\
\raisebox{1.7ex}[-1.7ex]{\mmxq}& 1.3 &    &  &  &  &  &  &  &  &  &  &  &     &             1.00 & 0.84 & 0.75 & 0.67\\
& 1.4 &     &        &     &       &     &     &      &      &       &      &      &      &     &  1.00 & 0.89 & 0.80\\
& 1.5 &     &        &     &       &     &     &      &      &       &      &      &       &     &     &  1.00 & 0.90\\
& 1.6 &     &        &     &       &     &     &      &      &       &      &      &      &      &      &      & 1.00\\ \hline\hline
\end{tabular}
}
\end{center}
\end{table*}

\end{document}